\documentclass{PoS}

\title{Light scalar mesons in 2+1 flavor full QCD 
       \thanks{This research is supported in part by the U.S. Department of
       Energy under grant DE-FG05-84ER40154.  The research of N.M. is supported
       in part by DST-SR/S2/RJN-19/2007 (India).  We thank CP-PACS/JLQCD for
       the use of their dynamical configurations.  We thank S. Prelovsek and
       A. Hasenfratz for important and interesting discussions.}  }
\ShortTitle{Light scalar mesons in 2+1 flavor full QCD}

\author{$\chi$QCD Collaboration: 
        \speaker{Terrence Draper}$^{a}$,
        Takumi Doi$^{a}$, Keh-Fei Liu$^{a}$, Devdatta Mankame$^{a}$, 
        Nilmani Mathur$^{b}$, Xiangfei Meng$^{a,c}$ \\
        {$^{a}$} Department of Physics and Astronomy, 
                 University of Kentucky, 
                 Lexington, KY 40506, USA \\
        {$^{b}$} Department of Theoretical Physics, 
                 Tata Institute of Fundamental Research,
                 Mumbai 40005, India \\
        {$^{c}$} College of Physics, 
                 Nankai University,
                 Tianjin 300071, P.R. China \\
        E-mail: \email{draper@pa.uky.edu}}

\abstract{We study the $a_{0}$ and $K_0^*$ light scalar mesons in 2+1 flavor
full QCD\@.  Particular attention is paid to fitting excited states, with an
eye toward determining whether scattering states are revealed.  An ultimate
goal will be to see how dynamical quarks affect the picture outlined with an
earlier quenched study using overlap fermions, namely, that it is the
$a_0(1450)$, not the $a_0(980)$, which is the lowest $\overline{q}q$ isovector
scalar state.  }

\FullConference{The XXVI International Symposium on Lattice Field Theory \\
                July 14 - 19, 2008\\
                Williamsburg, Virginia, USA}

\begin{document}

\section{Motivation}

\begin{figure}[ht]
  \vspace{0cm}
  \begin{center}
  \includegraphics[angle=0,width=0.5\hsize]{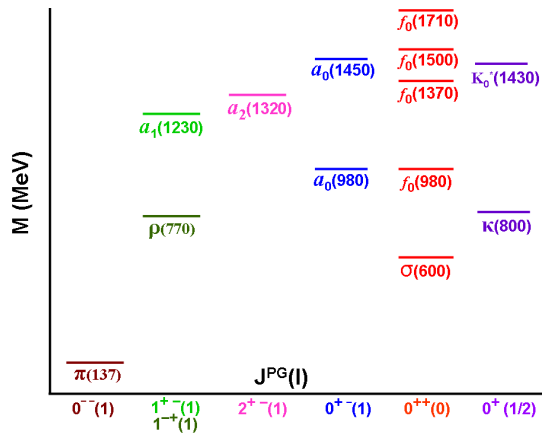}%
  \vspace{-0.1cm}
  \caption{\label{Fig:Expt} Spectrum of low-lying mesons.}
  \vspace{0cm}
  \end{center}
\end{figure}

The light scalar mesons are not unambiguously identified in terms of quark
content and their $SU(3)_f$ classification.  There are more experimental
candidates for the $\overline{q}q$ nonet than can be accommodated in the simple
quark model.  Fig.~\ref{Fig:Expt} shows the current experimentally
known-scalar, and other low-lying, mesons.

\begin{figure}[ht]
  \vspace{0cm}
  \begin{center}
  \includegraphics[angle=0,width=0.5\hsize]{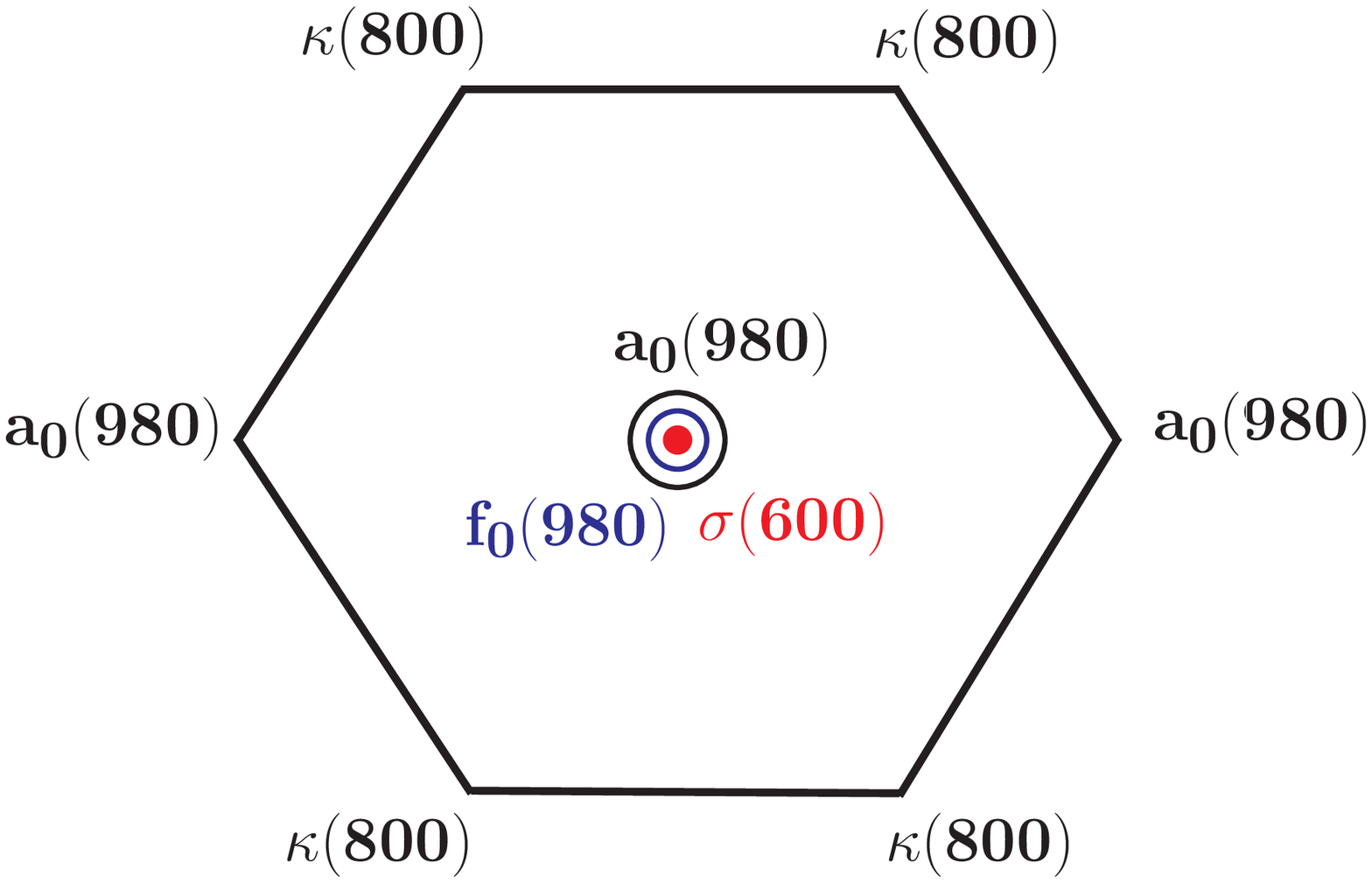}%
  \includegraphics[angle=0,width=0.5\hsize]{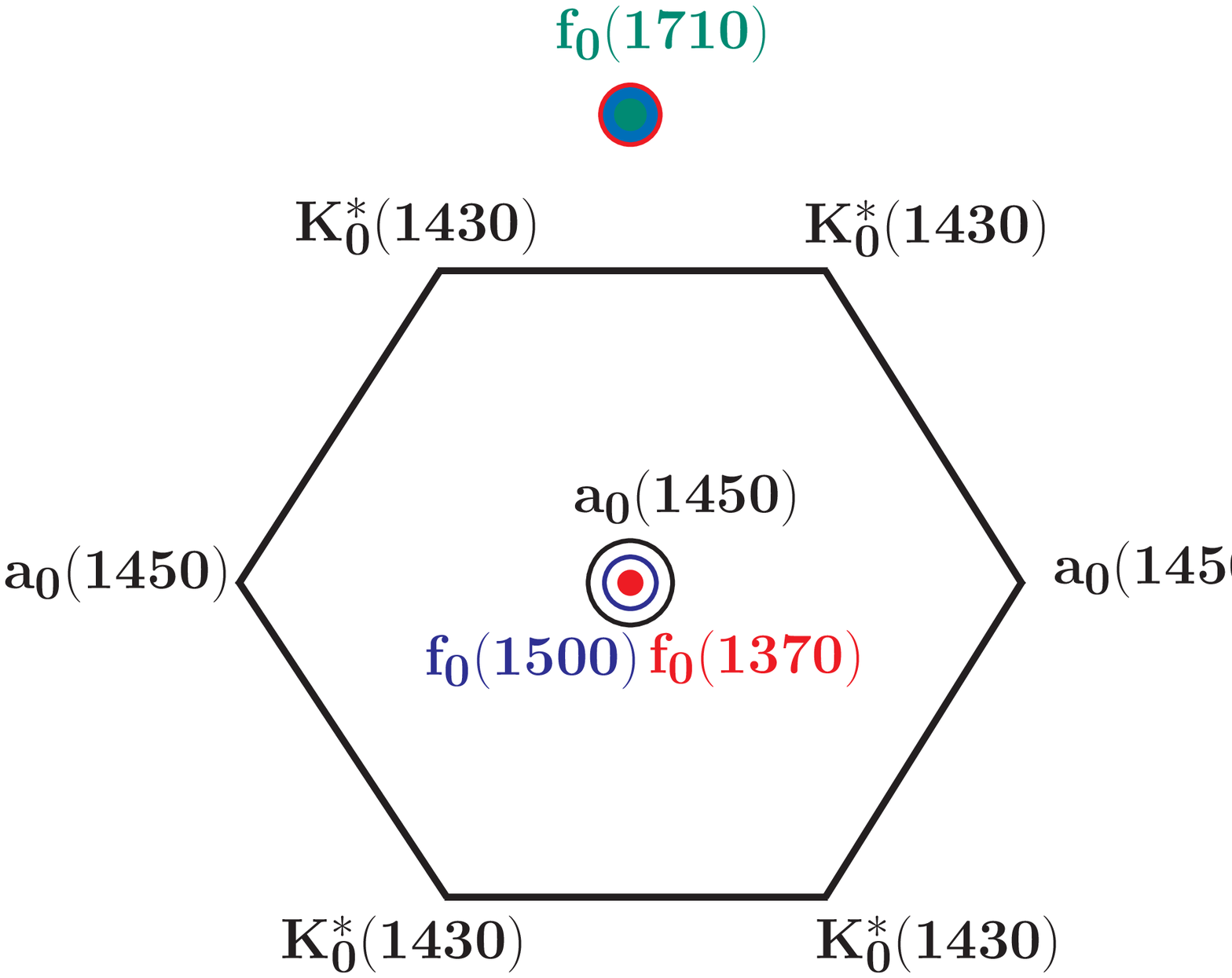}
  \vspace{-0.1cm}
  \caption{\label{Fig:Nonet} Proposed classification of light scalar mesons.}
  \vspace{0cm}
  \end{center}
\end{figure}

An attractive classification of light scalar mesons emerges~\cite{Liu07} which
resolves many puzzles in the ordering of states and size of branching ratios.
This is indicated in Fig.~\ref{Fig:Nonet} : a) a tetraquark nonet, including
the $\sigma(600)$, $a_0(980)$ and $f_0(980)$, below $1\,{\rm GeV}$, and b) a
$\overline{q}q$ nonet, including the $a_0(1450)$ and $K_0^*(1430)$, and an
almost pure glueball, $f_0(1710)$, above $1\,{\rm GeV}$.

Our quenched calculation with overlap fermions~\cite{Mathur07} offered support
of this view in providing evidence that the $\sigma(600)$ is a four-quark
state.  We also found evidence~\cite{Mathur07} that the lowest lying isovector
scalar $\overline{q}q$ state is the $a_0(1450)$, not the $a_0(980)$.

\pagebreak

\begin{figure}[ht]
  \vspace{0cm}
  \begin{center}
  \includegraphics[angle=0,width=0.5\hsize]{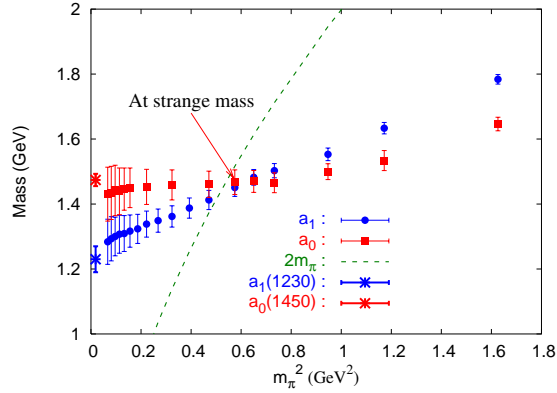}%
  \vspace{-0.1cm}
  \caption{\label{Fig:a0a1} $a_0$ and $a_1$ masses are plotted as a function of
          $m_{\pi}^2$. Also shown is the two pion mass (dashed line) which
          becomes lower than the $a_0$ around the strange quark mass region.}
  \vspace{0cm}
  \end{center}
\end{figure}

Fig.~\ref{Fig:a0a1} shows the masses of the $a_0$ and $a_1$ as a function of
quark mass.  The $a_0$ has little dependence on mass, especially below the
strange quark mass region, and extrapolates (from quite close to the chiral
limit with pion mass as low as $180\,{\rm MeV}$) to a mass of $1.42(13)\,{\rm
GeV}$.  Ghost would-be $\pi\eta'$ states, a quenched artifact, were identified
and handled by the fit, but no evidence of $a_0(980)$ was seen in this channel.
Indeed it was excluded with a squared spectral weight less than $0.015$ that of
the $a_0(1450)$.  This may account for why the $K_0^*(1430)$ and $a_0(1450)$
have approximately equal masses experimentally even though the former has a
strange quark.  This counters the conventional quark-counting rule; however,
the constituent quark model may have limitations for light hadrons where chiral
symmetry plays an important role.

The proposed picture, however intriguing, is nevertheless tentative.  Early
two-flavor dynamical simulations~\cite{Kunihiro04, Prelovsek04} reported values
at or above $1.5\,{\rm GeV}$, but later two-flavor dynamical
calculations~\cite{McNeile06, Frigori07, Hashimoto08} obtained a ground state
near $1\,{\rm GeV}$.  But is this the $a_0$ or $\pi\eta'_2$?  Dynamical
simulations must identify both the $a_0$ and scattering state(s) before the
issue is settled.

\section{Lattice Details}

We use the 2+1 flavor full QCD configurations provided by
CP-PACS+JLQCD~\cite{CP06}.  These have renormalization-group improved gauge
action and non-perturbatively $O(a)$-improved clover quark action.

The lattice size is $16^3\times 32$ with lattice spacing $a=0.12\,{\rm
fm}$. The light sea quark masses have $\kappa=0.13760$, $0.13800$, $0.13825$,
for which $m_{PS}(LL)/m_{V}(LL)= 0.71$, $0.66$, $0.62$, with pion masses
$0.84$, $0.70$, and $0.61\,{\rm GeV}$.  The strange sea quark mass has
$\kappa_S=0.13760$.  We use valence quark action and masses which match the sea
quark (i.e.\ no partial quenching).

We compute local-local two-point correlation functions with very high
statistics; for each of 800 configurations, we use 32 different delta-function
sources (requiring 32 different valence quark matrix inversions) well separated
in space and time.

We use the Sequential Empirical Bayes (SEB) method~\cite{Chen04}, a
constrained-curve fitting algorithm, to fit the ground and some excited states.

\section{Effective Masses and Fits}

We look first at the heaviest quark mass for which both strange and up/down
quark masses (for both valence and sea) are equal at $\kappa_S=\kappa=0.13760$
($m_\pi = 0.84\,{\rm GeV}$).

\begin{figure}[ht]
  \vspace{0cm}
  \begin{center}
  \includegraphics[angle=0,width=0.5\hsize]{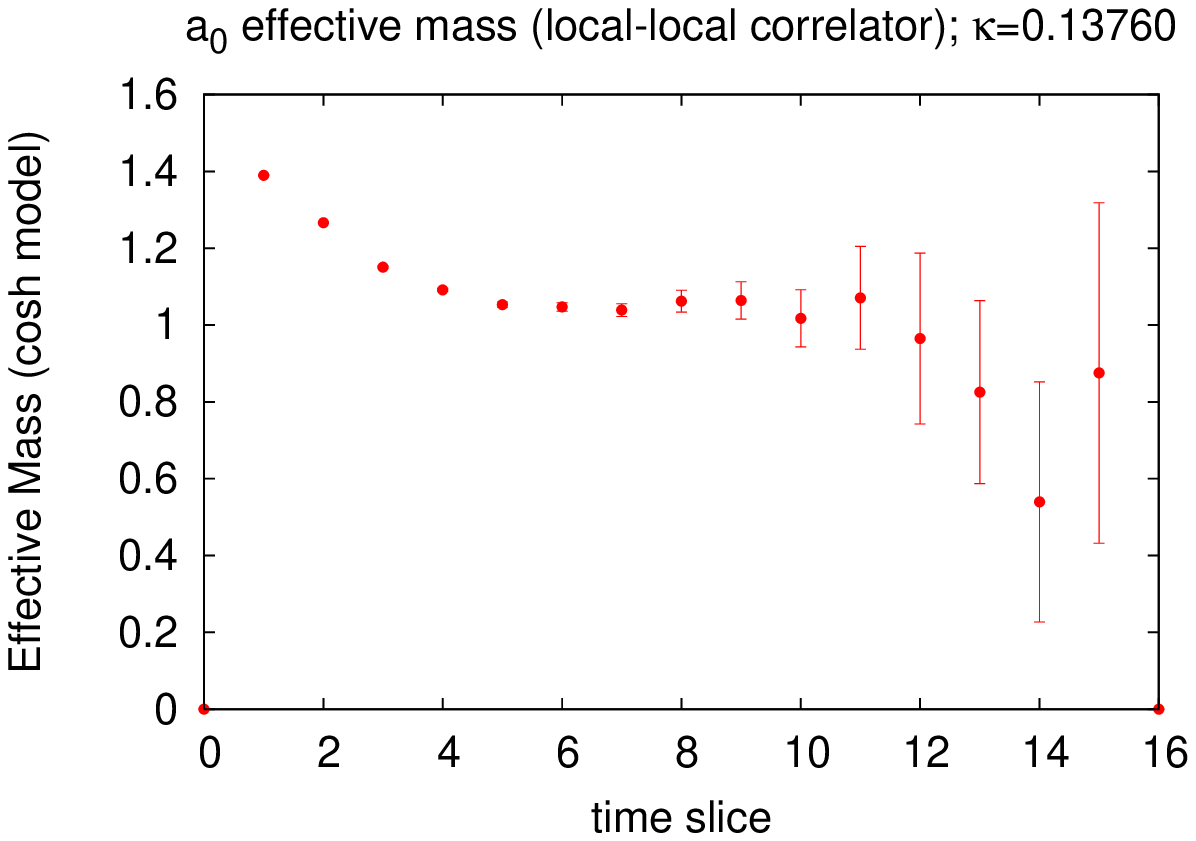}%
  \includegraphics[angle=0,width=0.5\hsize]{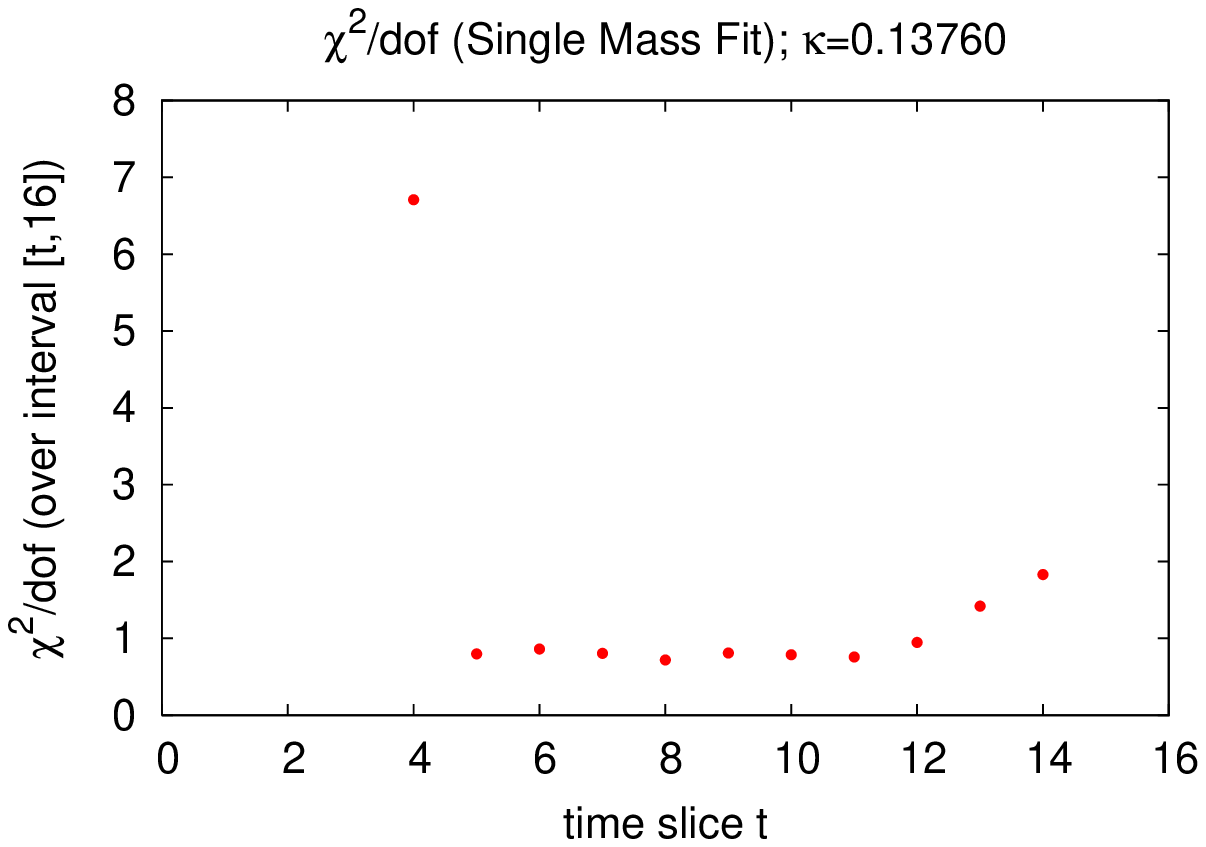}%
  \vspace{-0.1cm}
  \caption{\label{Fig:13760} Left: Effective mass plot for the isovector scalar
           channel with both strange and up/down quark masses (for both valence
           and sea) equal at $\kappa=\kappa_S=0.13760$ ($m_\pi = 0.84\,{\rm
           GeV}$). All effective mass plots are in accord with a cosh, not
           exponential, model; that is, for a correlation function (with
           periodic boundary conditions) which asymptotically is saturated with
           a single mass term, the effective mass plot will be a plateau right
           up to the middle of the lattice.  Right: $\chi^2/dof$ for one-state
           fits over the intervals [$t$,16]. }
  \vspace{0cm}
  \end{center}
\end{figure}

Fig.~\ref{Fig:13760} (left) shows the effective mass plot for the $a_0$ (which
is also the $K_0^*$ since these quark masses are degenerate).
Fig.~\ref{Fig:13760} (right) shows the $\chi^2/{\rm dof}$ for various one-state
(single-cosh) fits for the interval [$t$,16].  We get good one-state fits over
plateaus as long as [5,16].  Given a starting interval, the automated SEB
method~\cite{Ishii05} adds earlier and earlier time slices, and determines when
new terms need be added to the fit model.

Final single or multi-state fits give consistent results for the ground-state
mass; we obtain $1.04(1)$, i.e. $1.70(2)\,{\rm GeV}$, and a first-excited state
much higher.  The isovector scalar interpolating operator can excite $\pi\eta$
scattering states (as well as $\pi\eta'$ and $K\overline{K}$ states).  If $\pi$
and $\eta$ propagate in the same direction in time, then they give a
contribution with energy $E=m_\eta+m_\pi+E_{\rm int}$.  At this quark mass, the
scattering state $\pi\eta$ has an energy which is presumably close to that of
the $a_0$ (since $2m_\pi = 1.68\,{\rm GeV}$).  In this case, the SEB method
cannot resolve the two energies with our current statistics.

\begin{figure}[ht]
  \vspace{0cm}
  \begin{center}
  \includegraphics[angle=0,width=0.5\hsize]{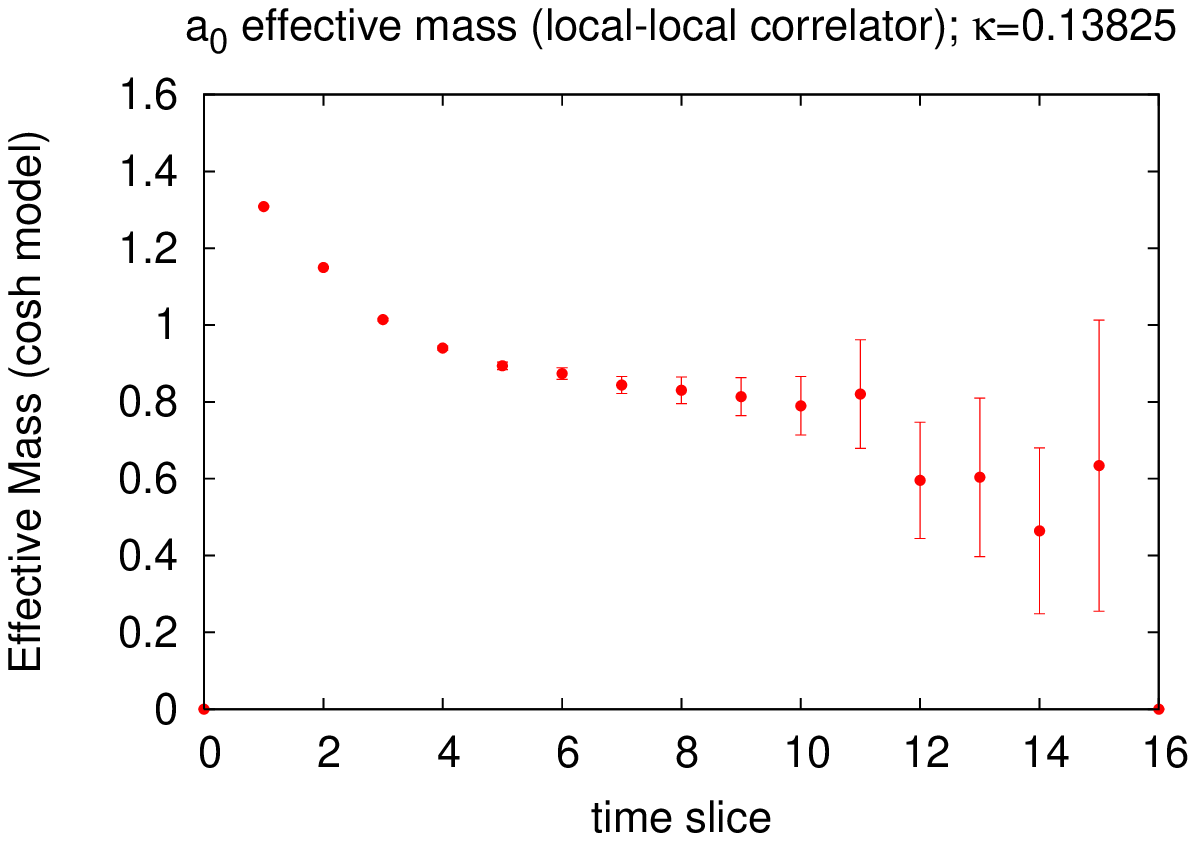}%
  \includegraphics[angle=0,width=0.5\hsize]{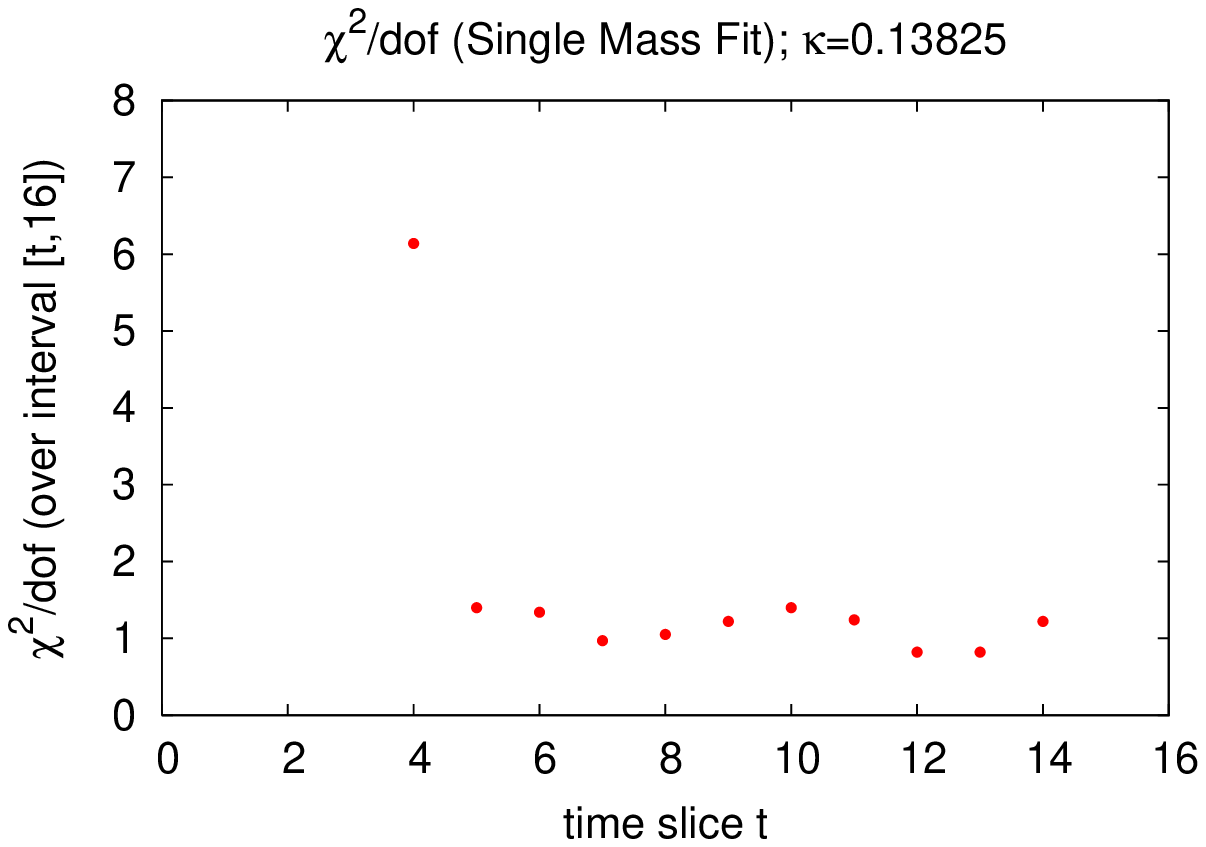}%
  \vspace{-0.1cm}
  \caption{\label{Fig:13825} Same as for Fig.~4 except that while the strange
           quark mass (for sea and valence) remains at $\kappa_S=0.13760$
           ($m_\pi = 0.84\,{\rm GeV}$), the light quark mass (for sea and
           valence) has $\kappa=0.13825$ ($m_\pi = 0.61\,{\rm GeV}$).}
  \vspace{0cm}
  \end{center}
\end{figure}

Next, we consider the case where the strange quark mass (for sea and valence)
remains at $\kappa_S=0.13760$ ($m_\pi = 0.84\,{\rm GeV}$), but the light quark
mass (for sea and valence) has $\kappa=0.13825$ ($m_\pi = 0.61\,{\rm GeV}$).
Fig.~\ref{Fig:13825} (left) shows the effective mass plot. Fig.~\ref{Fig:13825}
(right) shows the $\chi^{2}/{\rm dof}$ for various one-state (single-cosh) fits
for the interval [$t$,16].  Again we get low $\chi^2/{\rm dof}$ for a wide
range of values of $t$; that is, there is a long plateau.

Multi-state or single-state fits give consistent results for the ground-state
mass of $0.83(2)$, i.e. $1.35(3)\,{\rm GeV}$.  The first-excited state is much
higher.  Using the Gell-Mann-Okubo mass formula
$3m_{\eta}^{2}=4m_{K}^{2}-m_{\pi}^{2}$ and the measured kaon mass of
$0.68\,{\rm GeV}$ to estimate the mass of the $\eta$ as $m_\eta\approx
0.70\,{\rm GeV}$, we expect the $\pi\eta$ scattering state to have an energy
near $1.31\,{\rm GeV}$ which is close to the measured value of the ground state
energy.

The dip in the effective mass plot at large Euclidean time hints of evidence of
a peculiar effect due to a behavior of scattering states on a lattice with
periodic boundary conditions in time.  If the $\pi$ and $\eta$ propagate in
{\it opposite\/} directions in time, they make a contribution with energy
$\Delta m = m_\eta - m_\pi$~\cite{Prelovsek08}.  Indeed an
artificially-constructed correlation function with such a ``wrap-around''
contribution can agree with the data.  The dip is not statistically significant
however, and unfortunately, further doubling the statistics did not help.  We
cannot confirm this effect with our data, but it must be kept in mind in all
future simulations.  Rather than regard this as another contamination to
isolate and control in an effort to extract the $a_0$ mass, one should take the
optimistic view and regard this as a fascinating opportunity to measure the
$\eta$ mass!

In an attempt to better isolate the $a_0$ in the correlator by eliminating this
wrap-around effect, we recomputed the propagators with Dirichlet (fixed)
boundary conditions in time.  Unfortunately, our statistics are poor at large
Euclidean times, and we are unable to get an improved fit.

\begin{figure}[ht]
  \vspace{0cm}
  \begin{center}
  \includegraphics[angle=0,width=0.5\hsize]{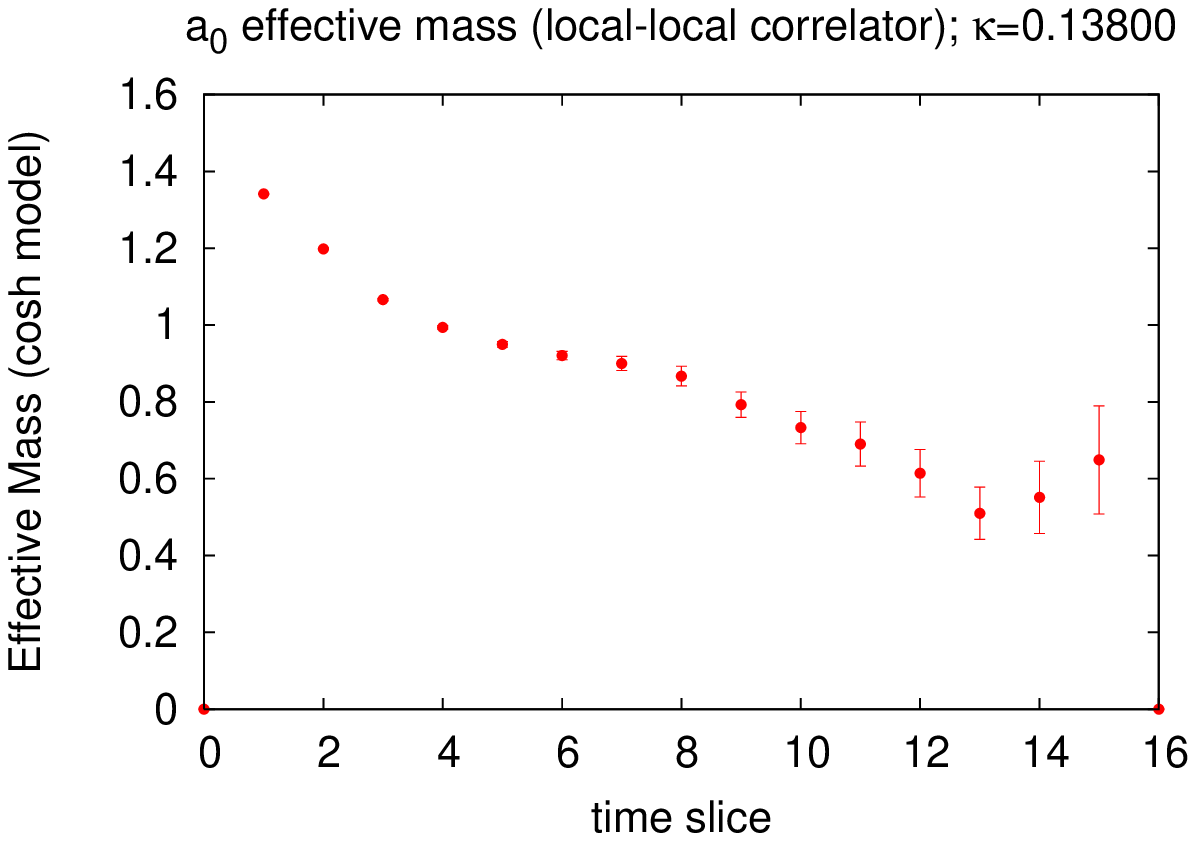}%
  \includegraphics[angle=0,width=0.5\hsize]{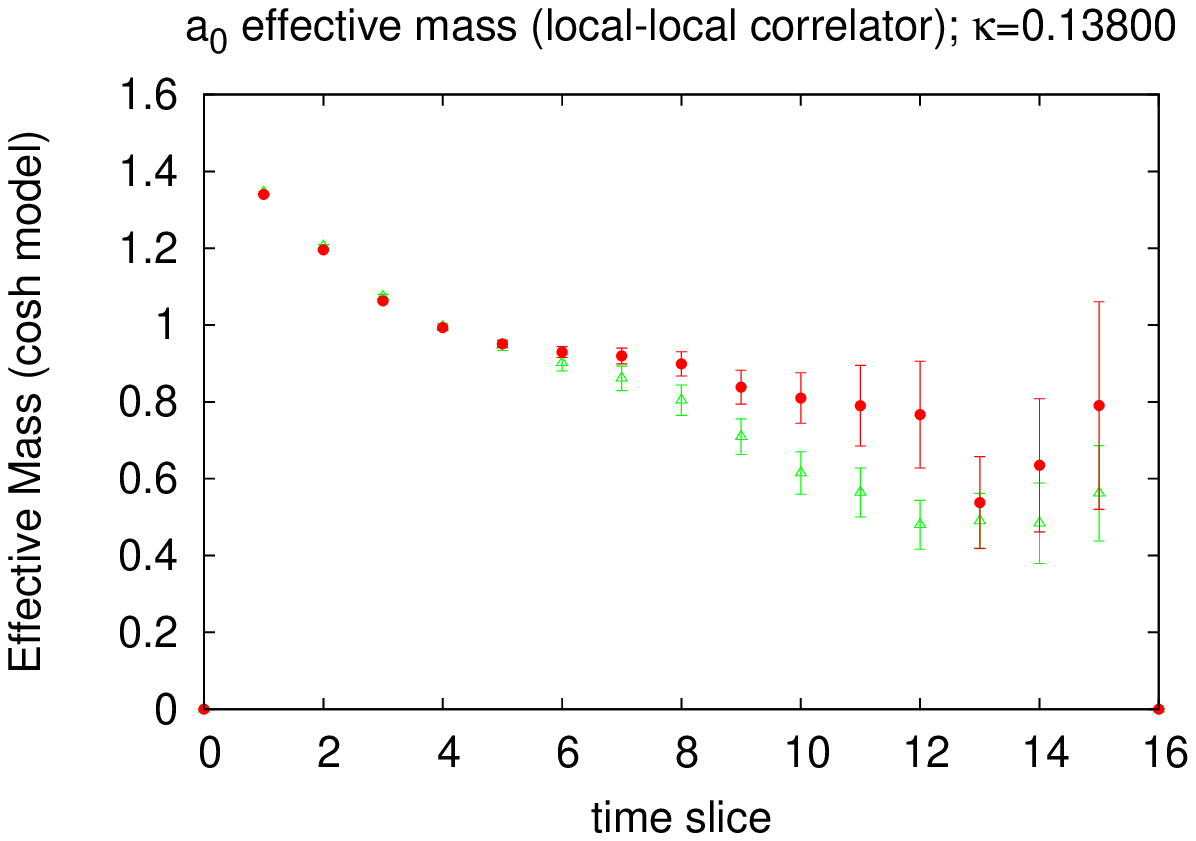}%
  \vspace{-0.1cm}
  \caption{\label{Fig:13800} Left: Same as for Fig.~4 (left) except that while
           the strange quark mass (for sea and valence) remains at
           $\kappa_S=0.13760$ ($m_\pi = 0.84\,{\rm GeV}$), the light quark mass
           (for sea and valence) has $\kappa=0.13800$ ($m_\pi = 0.61\,{\rm
           GeV}$). Right: The effective mass plot from the second trajectory of
           230 configurations (green triangles) and that of the remaining two
           trajectories combined (red dots).}
  \vspace{0cm}
  \end{center}
\end{figure}

The case where the strange quark mass (for sea and valence) remains at
$\kappa_S=0.13760$ ($m_\pi = 0.84\,{\rm GeV}$) and the light quark mass is
intermediate at $\kappa=0.13800$ ($m_\pi = 0.70\,{\rm GeV}$) is most peculiar.
Fig. 6 (left) shows the effective mass plot.
In contrast to the previous two cases, here there is no stable plateau, and the
error bars are smaller than expected.  We cannot make a fit in this case.

The dynamical gauge configurations were created from three trajectories.
Fig.~\ref{Fig:13800} (right) shows the effective mass plot from the second
trajectory of 230 configurations (green triangles) and that of the remaining
two trajectories combined (red dots).  The latter resembles those of the other
masses, but the former is peculiar with relatively small error bars and lack of
a plateau. This behavior of having one trajectory giving disparate results is
not replicated in other channels.

In summary, we are only able to obtain ground state fits in the isovector
scalar channel for two of the three light quark masses studied.  Although we
don't attempt a formal chiral extrapolation with only two points at rather
large quark mass (with $m_{\pi} = 0.84\,{\rm GeV}$ and $0.61\,{\rm GeV}$), the
downward trend suggests a ground state mass near $1\,{\rm GeV}$, rather than
near $1.5\,{\rm GeV}$, in this channel.  At face value, this suggests that the
lowest lying isovector scalar $\overline{q}q$ state is the $a_0(980)$, not the
$a_0(1450)$.  This agrees with that of other recent dynamical quark
simulations~\cite{McNeile06, Frigori07, Hashimoto08} and at first glance
appears to be at odds with the picture painted with a quenched overlap
calculation at low quark mass~\cite{Mathur07}.  But we cannot jump to this
conclusion because the measured ground state energies are so close to expected
$\pi\eta$ scattering states.  Dynamical simulations must identify both the
$a_0$ and scattering state(s) before the issue is settled.

The use of smeared sources would alter the proportion of $\overline{q}q$ and
scattering states, allowing SEB or a variational calculation to resolve the
states.  We might also expect a resolution with a new set of 2+1 flavor
dynamical gauge configurations at lighter quark masses.

\section{Hybrid Boundary Conditions}

For our quark masses, the kinematics are such that we expect there to be a
$\pi\eta$ scattering state near the measured ground state in the isovector
scalar channel.  To reveal this, we recomputed the valence quark propagators
with hybrid boundary conditions (HBC)~\cite{Ishii05}, using the same
configurations.  In our version of HBC, we use anti-periodic boundary
conditions in space for the valence quarks, while the dynamical quarks still
have periodic boundary conditions.  Although the valence quarks are given
non-zero momentum by the interpolating fields, a $\overline{q}q$ ground state
will still have zero momentum.  On the other hand, in the scattering state each
meson must have non-zero momentum (since the lowest Fock component has one
valence and one sea quark).  Thus, HBC raise the energy of the scattering
states while leaving that of $\overline{q}q$ unchanged, potentially allowing
SEB to resolve the states.  (A caveat is that with mixed boundary conditions
(valence versus sea), ghost states may complicate matters.)

Using HBC for $\kappa_S=0.13760$ and $\kappa=0.13825$, we measure the energy of
the lowest isovector scalar state to be $1.50(3)\,{\rm GeV}$.  This is higher
than the $1.35(3)\,{\rm GeV}$ measured with periodic boundary conditions (PBC).
This strongly suggests that the measured PBC ground state energy (GSE) is that
of a scattering state, and casts doubt that the $a_0$ mass extrapolates to a
value near $1\,{\rm GeV}$ in the chiral limit.  However, this HBC GSE does not
exceed the PBC GSE by the expected momentum-dependent amount, if both are
scattering states.  Rather, the measured HBC GSE is less, and could be that of
the $a_0$.

We propose the following scenario which is consistent with our (limited)
observations: at our lowest light quark mass the PBC GSE of $1.35(3)\,{\rm
GeV}$ is that of a scattering state, with the single-particle ($a_0$) state
lying higher.  As HBC are imposed, the scattering state energy increases,
leapfrogging that of the single-particle state which does not change much,
leaving the single particle state exposed as the (new) ground state at
$1.50(3)\,{\rm GeV}$.  Furthermore, at higher light quark mass (degenerate with
the strange), we observe that the GSE does not change much from PBC
($1.70(2)\,{\rm GeV}$) to HBC ($1.69(2)\,{\rm GeV}$), suggesting that the
ground state in this case is a single-particle ($a_0$) state.  In addition,
although a naive chiral extrapolation for PBC would give an extrapolated value
near $1\,{\rm GeV}$, this extrapolation should not be done, as it mixes
single-particle and scattering states!  The HBC GSE chiral extrapolation is
much less steep than for the PBC GSE, with an extrapolated value much higher
than $1\,{\rm GeV}$.

Unfortunately, our evidence for such a scenario is incomplete.  With the
current data set, the statistics are not sufficient for SEB to convincingly
extract excited states.  We need to measure these to verify that we see HBC
raise energies by the expected amounts.  Then we can unambiguously identify and
distinguish scattering states from single-particle states.  Until this is
resolved, we urge the community to keep an open mind about whether or not the
$a_0(980)$ is a $q\bar{q}$ state.

Scalar mesons are an increasingly rich, fascinating, and frustrating forum.

\end{document}